\documentclass[12pt,twoside]{article}
\usepackage{mathpazo} 
\usepackage[mathscr]{eucal}
\usepackage{amsmath,amsfonts,amssymb,amsthm}
\usepackage[matrix,arrow]{xy}
\usepackage{times}
\usepackage{pdfsync}
\usepackage{graphicx}
\usepackage{epstopdf}
\usepackage{dcolumn}
\usepackage{hyperref}
\usepackage[dvips]{color}
\usepackage[textmath]{preview}

\def\d_Vphi{\text{d}_V\hspace{-0.06em}\phi}
\def\d_Vphibar{\text{d}_V\hspace{-0.06em}\bar\phi}
\def\d_Vxi{\text{d}_V\hspace{-0.06em}\xi}

\def\be{\begin{eqnarray}}
\def\ee{\end{eqnarray}}
\def\beann{\begin{eqnarray*}}
\def\eeann{\end{eqnarray*}}
\def\beq{\begin{equation}}
\def\eeq{\end{equation}}
\def\ba{\begin{array}}
\def\ea{\end{array}}
\def\ben{\begin{enumerate}}
\def\een{\end{enumerate}}
\def\bea{\begin{eqnarray}}
\def\eea{\end{eqnarray}}

\def\5{\bar }
\def\6{\partial }
\def\7{\hat }
\def\4{\tilde }
\advance\textheight\topskip
\renewcommand{\tilde}{\widetilde}
\renewcommand{\hat}{\widehat}

\newcommand{\dd}{\partial}
\renewcommand{\d}{\partial}

\renewcommand{\geq}{\,{\geqslant}\,}

\newcommand{\binner}[2]{%
  {\langle}\kern-4.15pt{\langle}#1{,}\,#2{\rangle}\kern-4.15pt{\rangle}}

\newcommand{\half}{\frac{1}{2}}

\newcommand{\ffrac}[2]{\raisebox{.5pt}%
  {\footnotesize$\displaystyle\frac{#1}{#2}$}\kern1pt}

\newcommand{\ddl}[2]{\ffrac{\dd #1}{\dd #2}}

\def\cH{\mathcal{H}}

\def\cL{\mathcal{L}}

\def\cP{\mathcal{P}}

\numberwithin{equation}{section} \makeatletter
\@addtoreset{equation}{section}

\DeclareFontFamily{OT1}{rsfs}{} \DeclareFontShape{OT1}{rsfs}{m}{n}{
<-7> rsfs5 <7-10> rsfs7 <10-> rsfs10}{}
\DeclareMathAlphabet{\mycal}{OT1}{rsfs}{m}{n}

\begin{document}

\def\mytitle{
Three-dimensional Bondi-Metzner-Sachs invariant two-dimensional field
    theories as the flat limit of Liouville theory
}

\pagestyle{myheadings} \markboth{\textsc{\small G.~Barnich,
    A.~Gomberoff, H.~Gonz\'alez}}{%
  \textsc{\small $\mbox{BMS}_{3}$ invariant Liouville theories}} 
\addtolength{\headsep}{4pt}

\begin{flushright}\small
ULB-TH/12-17\end{flushright}

\begin{centering}

  \vspace{1cm}

  \textbf{\Large{\mytitle}}


  \vspace{1.5cm}

  {\large Glenn Barnich$^{a}$}

\vspace{.5cm}

\begin{minipage}{.9\textwidth}\small \it  \begin{center}
   Physique Th\'eorique et Math\'ematique \\ Universit\'e Libre de
   Bruxelles and International Solvay Institutes \\ Campus
   Plaine C.P. 231, B-1050 Bruxelles, Belgium
 \end{center}
\end{minipage}

\vspace{.5cm}

 {\large Andr\'es Gomberoff}

\vspace{.5cm}

\begin{minipage}{.9\textwidth}\small \it  \begin{center}
   Departamento de Ciencias F\'{\i}sicas,  Universidad Andres Bello \\
 Av. Rep\'{u}blica 252, Santiago, Chile
\end{center}
\end{minipage}

\vspace{.5cm}

{\large Hern\'an A. Gonz\'alez}

\vspace{.5cm}

\begin{minipage}{.9\textwidth}\small \it \begin{center}
   Departamento de F\'{\i}sica, P. Universidad Cat\'olica de
Chile \\ Casilla 306, Santiago 22, Chile
\end{center}
\end{minipage}

\end{centering}

\vspace{1cm}

\begin{center}
  \begin{minipage}{.9\textwidth}
    \textsc{Abstract}. In the gravitational context, Liouville theory
    is the two-dimensional conformal field theory that controls the
    boundary dynamics of asymptotically \({\rm AdS}_3\) spacetimes at
    the classical level. By taking a suitable limit of the coupling
    constants of the Hamiltonian formulation of Liouville, we
    construct and analyze a \({\rm BMS}_3\) invariant two-dimensional
    field theory that is likely to control the boundary dynamics at
    null infinity of three dimensional asymptotically flat gravity.
  \end{minipage}
\end{center}

\vfill

\noindent
\mbox{}
\raisebox{-3\baselineskip}{%
  \parbox{\textwidth}{\mbox{}\hrulefill\\[-4pt]}}
{\scriptsize$^a$Research Director of the Fund for Scientific
  Research-FNRS Belgium. E-mail: gbarnich@ulb.ac.be}

\thispagestyle{empty}
\newpage


\section{Introduction}
\label{sec:introduction-1}

As a non trivial two-dimensional conformal field theory, Liouville
theory is ubiquitous in theoretical physics (see
e.g.~\cite{Seiberg:1990eb,Teschner:2001rv,Nakayama:2004vk} for
reviews). In particular, in the context of three dimensional
asymptotically anti-de Sitter spacetimes, and more generally the
AdS/CFT correspondence (see e.g.~section 5.5 of
\cite{Aharony:1999ti}), Liouville theory controls the boundary
dynamics \cite{Coussaert:1995zp} on the classical level : starting
from the Chern-Simons formulation of anti-de Sitter gravity
\cite{Achucarro:1986vz,Witten:1988hc}, it is obtained through a
Hamiltonian reduction from a suitable Wess-Zumino-Witten model by
taking into account gravitational boundary conditions.

For flat three dimensional gravity, asymptotic dynamics that is as
rich as the one of the anti-de Sitter case can be defined at null
infinity \cite{PhysRevD.55.669,Barnich:2006avcorr,Barnich:2010eb}. It
can be connected through a well-defined flat-space limit to the
anti-de Sitter case \cite{Barnich:2012aw}: the limit of the BTZ black
holes are cosmological solutions whose horizon entropy can be
understood from symmetry arguments \cite{Barnich:2012he,Bagchi:2012he}
consistent with those of the anti-de Sitter case
\cite{Strominger:1998eq}.

In this context of flat space holography, a natural problem is to
construct the action that controls the boundary dynamics by starting
from the Chern-Simons formulation of flat gravity and taking the
gravitational boundary conditions into account. This will be addressed
in detail elsewhere.

In this note, we take a short-cut and directly construct a candidate
for such an action: by taking appropriate ``flat'' limits of Liouville
theory, we construct two ${\rm BMS}_3$ invariant two-dimensional field
theories and work out their Poisson algebra of conserved
charges. Whereas the first limit has no central extension, the second
one admits a central extension of exactly the same type than in the
gravitational surface charge algebra.

The constructed theories are interacting two-dimensional field
theories with a symmetry group that is of the same dimension than the
conformal algebra. We briefly elaborate on some of their classical
properties by working out the anomalous transformations laws of their
energy momentum tensors on-shell and relating them to the general
solution of the field equations obtained from a suitable free field.

\section{Liouville theory, flat limits and \(\mbox{BMS}_3\) invariance}
\label{sec:liouv-theory-flat}

We start by writing the Liouville action in Hamiltonian form on the
Minkowskian cylinder with time coordinate time\footnote{The choice of
  the letter \(u\) for time is due to the fact that the time of the
  flat limit is a null time in the gravitational context.}  \(u\),
angular coordinate \(\phi\in [0,2\pi)\) and metric
\(\eta_{\mu\nu}={\rm diag}(-1,l^2)\),
\begin{equation}
\label{hamiltonianLiouville}
I_H[\varphi,\pi;\gamma,\mu,l]=\int du d\phi\, \cL_H ,\quad 
\cL_H=\pi \dot{\varphi}-
\frac{1}{2}\pi^2-\frac{1}{2l^2}{\varphi'}^2-\frac{\mu}{2\gamma^2} 
e^{\gamma\varphi}.
\end{equation}
In this parametrisation, if \(L\) is the basic physical dimension of
length, \([\varphi]=L^\half\), \([\pi]=L^{-\half}\),
\([\gamma]=L^{-\half}\), \([l]=L\), and \([\mu]=[L]^{-2}\). The
cylinder coordinates are related to the light-cone variables through
\(x^\pm=\frac{u}{l}\pm \phi\). Under
two-dimensional conformal transformations \(\tilde x^+=F(x^+)\),
\(\tilde x^-=G(x^-)\), the Lagrangian action is invariant if the field
transforms as 
\begin{equation}
  \label{eq:18}
  \tilde \varphi(\tilde x)=\varphi(x)-\frac{1}{\gamma}\ln{
    F'G'}.
\end{equation}
This invariance is lifted to the Hamiltonian action through
\begin{multline}
  \tilde \pi(\tilde x)=\frac{1}{\sqrt{F'G'}}\left(\pi(x)-
\frac{1}{l}(\d_++\d_-)\varphi (x) \right)+\\
+\frac{1}{l}\left(\frac{1}{F' 
      }\d_++\frac{1}{G'}\d_-\right)\left(\varphi(x)-\frac{1}{\gamma}\ln{F'G'}
\right).
\end{multline}

We are interested in two types of ``flat'' limits of the Hamiltonian
Liouville action. The first consists in just taking \(l\to \infty\)
with \(\gamma,\mu\) fixed,
\begin{equation}
  \label{eq:20}
I_H[\varphi,\pi;\gamma,\mu]=\int du d\phi\, \cL_H ,\quad
\cL_H=\pi 
\dot{\varphi}-
\frac{1}{2}\pi^2-\frac{\mu}{2\gamma^2} e^{\gamma\varphi}.
\end{equation}
In this case, it is still possible to eliminate the momentum by its
 equation of motion leading to 
 \begin{equation}
   \label{eq:6}
   I_L[\varphi;\gamma,\mu]=\int du d\phi
   \left(\frac{1}{2}\dot{\varphi}^2-
\frac{\mu}{2\gamma^2} e^{\gamma\varphi}\right).
 \end{equation}

For the second limit, we first rescale the field and its
momentum through a canonical transformation,
\begin{equation}
\varphi = l\Phi, \quad \pi=\frac{\Pi}{l}, 
\end{equation}
and then take the limit
while keeping
\(\beta=\gamma l, \nu= \mu l^2\) fixed
so that 
\begin{equation}
  \label{eq:21}
 I_H[\Phi,\Pi;\beta,\nu]=\int du d\phi\, \cL_H ,
\quad  \cL_H=\Pi \dot{\Phi}-
\frac{1}{2}\Phi'^2-\frac{\nu}{2\beta^2} e^{\beta\Phi}. 
\end{equation}
Even though there is no local second order version of this theory, one
can in principle eliminate \(\Phi\) from the action. In order to so,
one has to solve the equations of motion of \(\Phi\) in terms of
\(\dot \Pi\) at the price of sacrificing spatial locality. In this
way, one ends up with a theory for \(\Pi\) that is of second order in
time derivatives. For example, in the mini-superspace approximation
where the canonical fields do not depend on \(\phi\), one ends up with 
\begin{equation}
  \label{eq:39}
  \cL_L=-\frac{\dot\Pi}{\beta}\ln{\dot{|\Pi|}} . 
\end{equation}

The \({\rm BMS}_3\) group admits a realization in terms of coordinate
transformations of \(S^1\times\mathbb R\) of the form
  \begin{equation}
    \label{eq:bms3group}
    \tilde\phi=\tilde\phi(\phi),\quad \tilde u=
    \tilde\phi'(u+\alpha(\phi)),\ 
  \end{equation}
  where the tensor density $\alpha$ transforms as
  \(\tilde\alpha(\tilde\phi)=(\alpha\tilde\phi')(\phi)\).
  It is then straightforward to check that action \eqref{eq:20} is
  invariant under
\begin{equation}
\tilde\varphi(\tilde u,\tilde\phi)=\varphi(u,\phi)-
\frac{2}{\gamma}\ln{|\tilde\phi'|},\quad 
\tilde\pi(\tilde u,\tilde \phi)=\frac{1}{\tilde \phi'}\pi(u,\phi),\label{eq:14}
\end{equation}
while action \eqref{eq:21} is invariant under 
\begin{equation}
  \label{eq:22}
\begin{split}
  & \tilde\Phi(\tilde u,\tilde\phi)=\Phi(u,\phi)-
  \frac{2}{\beta}\ln{|\tilde\phi'|},\\
  & \tilde\Pi(\tilde u,\tilde \phi)=\frac{1}{\tilde \phi'}
  \Pi(u,\phi) +\half\tilde\phi' (\ddl{u}{\tilde\phi})^2\partial_u
  \Phi+\ddl{u}{\tilde\phi}\d_\phi
  \Phi-\frac{2}{\beta(\tilde\phi')^2}\Big(\tilde
        u''-2 \frac{\tilde u'}{\tilde\phi'}\tilde\phi''\Big),
\end{split}
\end{equation} 
by using \(\ddl{\tilde u}{u}=\ddl{\tilde\phi}{\phi}\),
\(\ddl{\tilde\phi}{u}=0\), \(\ddl{u}{\tilde
  u}=\ddl{\phi}{\tilde\phi}\), \(\ddl{\phi}{\tilde u}=0\),
\(\ddl{u}{\tilde\phi}=-\frac{\tilde u'}{(\tilde\phi')^2}\) and also
  \(\frac{(\tilde \phi'')^2}{(\tilde\phi')^2}=\d_u (\frac{\tilde
    u'\tilde\phi''}{(\tilde\phi')^2})\).

\section{Poisson algebra of conserved charges}
\label{sec:poiss-algebra-cons}

\subsection{Liouville theory}
\label{sec:liouville-theory}

In the current set-up, if
\(\xi=f\d_u+Y\d_\phi\) is a conformal Killing vector on the cylinder
\begin{equation}
\partial_{u}f=\partial_{\phi}Y,
\quad \partial_{u}Y=\frac{1}{l^2}\partial_{\phi}f, \label{eq:1}
\end{equation}
or, equivalently, \(f=\frac{l}{2} (Y^++Y^-)\) with \( Y=\half
(Y^+-Y^-)\), \(Y^+=Y^+(x^+), Y^-=Y^-(x^-)\), the infinitesimal
symmetry transformations of the field and its momentum are given by
 \begin{equation}
\label{symLio1}
\begin{split}
  -\delta_\xi \varphi&=f \pi+Y\varphi'+\frac{2}{\gamma} Y',\\
  -\delta_\xi \pi&=-f
  \frac{\mu}{2\gamma}e^{\gamma{\varphi}}+(\frac{1}{l^2}f
  \varphi')' +(\pi Y)'+\frac{2}{\gamma l^2} f''.
\end{split}
\end{equation}
They are related to the infinitesimal versions of the finite
transformations discussed in the previous section through trivial
equations-of-motion symmetries chosen so as to remove the
time-derivatives of the canonical variables.

Invariance of the action follows from 
\begin{multline}
  -\delta_\xi \cL_H =\partial_{\phi}\Big(Y\big[\pi\dot{\varphi}-
  \frac{1}{2}\pi^2-\frac{1}{2l^2}{\varphi'}^2-  \frac{\mu}{2\gamma^2}
e^{\gamma\varphi}\big]-\frac{2}{\gamma l^2}Y'' \varphi
  +\frac{1}{l^2} f(\dot{\varphi} -\pi) \varphi'\Big) \\+
\partial_u\Big(f\big[\frac{1}{2}\pi^2-\frac{1}{2l^2}{\varphi'}^2-
\frac{\mu}{2\gamma^2}
        e^{\gamma\varphi}\big]+\frac{2}{\gamma l^2} f''
      \varphi \Big).
\end{multline}
Writing \(-\delta_\xi\cL_H=\d_\mu k^\mu_\xi\), the canonical Noether
current is given by \(-j^\mu_\xi=\ddl{\cL_H}{\d_\mu \varphi}\delta_\xi
\varphi+\ddl{\cL_H}{\d_\mu \pi}\delta_\xi \pi+k^\mu_\xi\), or
explicitly
\begin{equation}
  \label{eq:31}
  \begin{split}
    j^u_\xi & =f\Big(\half \pi^2+\frac{1}{2l^2} {\varphi'}^2
    +\frac{\mu}{2\gamma^2} e^{\gamma\varphi}\Big) -\frac{2}{\gamma l^2} f''
    \varphi
    + Y\pi \varphi' +\frac{2}{\gamma} Y' \pi ,\\
    j^\phi_\xi & =-\frac{1}{l^2} f\pi\varphi'-Y\left(\half \pi^2
    +\frac{1}{2l^2} {\varphi'}^2 -\frac{\mu}{2\gamma^2}
    e^{\gamma\varphi}\right)-\frac{2}{\gamma l^2} Y' \varphi'
    +\frac{2}{\gamma l^2} Y'' \varphi,
  \end{split}
\end{equation}
where the equations of motion have been used to eliminate time
derivatives in the spatial part of the Noether current. Defining
\(j^\mu_\xi=-{T^\mu}_\nu \xi^\nu+\d_\nu k_\xi^{[\nu\mu]}\) with
\(k^{[\phi u]}_\xi=-\frac{2}{\gamma l^2} f'\varphi+\frac{2}{\gamma l^2}
f\varphi' +\frac{2}{\gamma} Y \pi\) and using again equations of
motions to eliminate time-derivatives gives the symmetric and
traceless energy-momentum tensor with components
\(T_{uu}=\mathcal{H}=\frac{1}{l^2} T_{\phi\phi}\), \(T_{u\phi}=
\mathcal{P}\) where 
\begin{equation}
  \label{eq:32}
\mathcal{H}= 
\half \pi^2 +\frac{1}{2l^2} {\varphi'}^2 +\frac{\mu}{2\gamma^2}
e^{\gamma \varphi} -\frac{2}{\gamma l^2} \varphi'',\quad 
\mathcal{P}=\pi \varphi' -\frac{2}{\gamma} \pi'.
\end{equation}

The associated Noether charge \(Q_\xi=\int_0^{2\pi} d\phi
j^u_\xi\) is
\begin{equation}
\label{NC}
  Q_\xi=\int^{2\pi}_0 d\phi\left[f\mathcal{H}+Y\mathcal{P} \right].
\end{equation}

In terms of the canonical equal-time Poisson bracket,
\(\{\varphi(u,\phi_1),\pi(u,\phi_2)\}=\delta(\phi_1-\phi_2)\),
 the charges generate the symmetry transformations
\eqref{symLio1} through \(-\delta_\xi z^a =\{z^a,Q_\xi\}\), and the
algebra of their integrands is
\begin{equation}
  \label{eq:2}
  \{Q_{\xi_1},Q_{\xi_2}\}=Q_{[\xi_1,\xi_2]_H}+K_{\xi_1,\xi_2},
\end{equation}
where 
\begin{equation}
  \label{eq:3}
  \hat f=f_1 Y'_2+Y_1f_2'-(1\leftrightarrow 2),\quad \hat
  Y=\frac{1}{l^2} f_1 f'_2 +Y_1Y_2'-(1\leftrightarrow 2),\\
\end{equation}
and the central extension is
\begin{equation}
  \label{eq:4}
  K_{\xi_1,\xi_2}=\frac{4}{\gamma^2 l^2}\int_0^{2\pi}d\phi \big[f'_1 
  Y''_2-(1\leftrightarrow 2)\big]. 
\end{equation}
The bracket $[\xi_1,\xi_2]_H=\hat f\d_u+\hat Y\d_\phi$ is related to
the standard Lie bracket by eliminating the time derivatives using the
conformal Killing equation \eqref{eq:1}. Algebra \eqref{eq:2} implies
in particular that the charges are conserved. Indeed, \(H=Q_{\d_{u}}\)
and conservation means that \(\ddl{}{u} Q_\xi+\{Q_\xi,H\}=0\). This is
encoded in \eqref{eq:2} by choosing \(\xi_1=\xi,\xi_2=\d_{u}\).

In terms of Fourier modes, the conformal Killing vectors of the
cylinder are given by 
\begin{equation}
  \label{eq:13}
\begin{split}
  p_m=e^{im\phi}\frac{1}{2l}\big[(e^{im\frac{u}{l}}+e^{-im\frac{u}{l}})l\d_u+
(e^{im\frac{u}{l}}-e^{-im\frac{u}{l}})\d_\phi\big],\\
j_m=e^{im\phi}\frac{1}{2}\big[(e^{im\frac{u}{l}}-e^{-im\frac{u}{l}})l\d_u+
(e^{im\frac{u}{l}}+e^{-im\frac{u}{l}})\d_\phi\big].
\end{split}
\end{equation}
If we denote the associated charges by 
\begin{equation}
  P_m = Q_{p_m},\quad J_m =
  Q_{j_m}\label{eq:11},
\end{equation}
their algebra reads
\begin{equation}
\label{VirPJ}
\begin{split}
&i\{P_{m},P_{n}\}=\frac{1}{l^2}(m-n)J_{m+n},\\
&i\{J_{m},J_{n}\}=(m-n)J_{m+n},\\
&i\{J_{m},P_{n}\}=(m-n)P_{m+n}+
\frac{8\pi}{\gamma^2l^2}m^3\delta_{m+n}.
\end{split}
\end{equation}
The change of basis \(P_m=l^{-1}(L^{+}_{m}+L^{-}_{-m})\) and
\(J_{m}=L^{+}_{m}-L^{-}_{-m}\) transforms this algebra into two copies
of the Virasoro algebra, \(i\{L^\pm_m,L^\pm_n\}=(m-n)
L^\pm_{m+n}+\frac{c^\pm}{12} m^3\delta_{m+n}\),
\(i\{L^\pm_m,L^\mp_n\}=0\) with \(c^{\pm}=
\frac{48\pi}{\gamma^2l}\). This is consistent with the Dirac bracket
algebra of surface charges in three dimensional asymptotically anti-de
Sitter spacetimes, normalized with respect to the \(M=0=J\) BTZ black
hole, which has central charges \(c^\pm=\frac{3l}{2G}\)
\cite{Brown:1986nw}.  If one uses the normalization of the action as
is given in Eq. (\ref{hamiltonianLiouville}), the theory is equivalent
to (2+1)-dimensional gravity \cite{Coussaert:1995zp}, when its
coupling constants are related to the gravitational ones by,
\begin{equation}
\begin{split}
\label{couplings}
G =\frac{\gamma^2l^2}{32\pi}, \  \ \Lambda=-\frac{1}{l^2}.
\end{split}
\end{equation}
where $\Lambda$ is the cosmological constant and $G$ is Newton's
constant.

Written in terms of these parameters, this is precisely the
Brown-Henneaux central charge, 
\begin{equation}
\begin{split}
\label{BrownHenneaux}
c^{\pm} = \frac{48\pi}{\gamma^2l}= \frac{3l}{2G}.
\end{split}
\end{equation}

\subsection{Gravitational results for 3d asymptotically flat spacetimes}
\label{sec:grav-results-three}

The Dirac bracket algebra of surface charges for asymptotically flat
three dimensional spacetimes at null infinity
\cite{Barnich:2006avcorr,Barnich:2010eb}, normalized with respect to
the null orbifold which is defined to have zero mass\footnote{See also
  \cite{Barnich:2012aw}, where the algebra it is normalized with
  respect to global Minkowski space. This amounts to shift $P_0$ by
  $-c_2/12$.}, is the centrally extended \(\mathfrak{bms}_3\) algebra
\begin{equation}
\begin{split}
\label{bms3}
&i\{P_{m},P_{n}\}=0,\\
&i\{J_{m},J_{n}\}=(m-n)J_{m+n}+\frac{c_1}{12}m^3\delta_{m+n},\\
&i\{J_{m},P_{n}\}=(m-n)P_{m+n}+\frac{c_2}{12}m^3\delta_{m+n}.
\end{split}
\end{equation}
with gravitational values \(c_1=0\), \(c_2=\frac{3}{G}\).

\subsection{Non-centrally extended limit}
\label{sec:non-centr-extend}

The first limit $l\to\infty$ simply amounts to dropping all terms
involving \(l^{-2}\) in formulas \eqref{eq:1}-\eqref{VirPJ}, with the
exception of \eqref{eq:13}. In particular, the general solution to
\eqref{eq:1} for $l\to\infty$ is given by
\begin{equation}
  \label{eq:7}
  f=T(\phi)+uY',\quad Y=Y(\phi), 
\end{equation}
for arbitrary functions \(T,Y\) of \(\phi\). The transformations
simplify to 
\begin{equation}
\label{symLio2}
\begin{split}
  -\delta_\xi \varphi&=f \pi+Y\varphi'+\frac{2}{\gamma} Y',\\
  -\delta_\xi \pi&=-f
  \frac{\mu}{2\gamma}e^{\gamma{\varphi}} +(\pi Y)',
\end{split}
\end{equation} 
while the Hamiltonian density in the expression for the Noether charge
\eqref{NC} reduces to
\begin{equation}
  \label{eq:8}
  \cH=\frac{1}{2}\pi^2+\frac{\mu}{2\gamma^2} e^{\gamma\varphi}, 
\end{equation}
with \(\cP\) unchanged.  At the same time, the components of the
energy-momentum tensor are given by
\(-{T^u}_u=\mathcal{H}={T^\phi}_\phi\), \(-{T^u}_\phi=\mathcal P\), \(
{T^\phi}_u=0\).  In the algebra, the central extension
\(K_{\xi_1,\xi_2}\) vanishes while \eqref{eq:3}, re-written in terms
of \((T,Y)\), turns into
\begin{equation}
  \label{eq:3a}
  \hat T=T_1 Y'_2+Y_1T_2'-(1\leftrightarrow 2),\quad \hat
  Y=Y_1Y_2'-(1\leftrightarrow 2).\\
\end{equation}
In terms of modes, which now become  
\begin{equation}
P_m=Q_{e^{im\phi}\d_u},\quad 
J_m=Q_{e^{im\phi}(im u\d_u+\d_\phi)},\label{eq:10}
\end{equation}
one then finds \eqref{bms3} with $c_1=0=c_2$. In other words, while
the theory defined by \eqref{eq:20} is invariant under
$\mathfrak{bms}_3$ transformations, the associated Poisson algebra of
Noether charges has no central extension. Hence this theory is not
related to asymptotically flat gravity in three dimensions, which is
known to have a non vanishing central extension in its corresponding
algebra, as we discussed in Section 3.2.  Another way of looking at
this is to notice that this limit of vanishing cosmological constant
produces $G \to \infty$. This can be seen from \eqref{couplings} when
keeping $\gamma$ fixed as $l\rightarrow\infty$.

\subsection{Centrally extended limit}
\label{sec:centr-extend-limit}

Since the rescaling of variables is a canonical transformation, the
Poisson algebra \eqref{eq:2}, or equivalently \eqref{VirPJ}, is
unchanged before taking the limit. After redefining the constants and
then taking the limit, the symmetry
transformations reduce to
\begin{equation}
-\delta_\xi \Phi=Y\Phi'+\frac{2}{\beta} Y',\quad
-\delta_\xi \Pi=-f\frac{\nu}{2\beta}
e^{\beta\Phi}+(f\Phi')'+(\Pi Y)'+\frac{2}{\beta} f'',
\end{equation}
where now \(\partial_{u}f=\partial_{\phi}Y\) and \(\partial_{u} Y=0\),
or equivalently, \eqref{eq:7} holds. Their generators can be written
as in \eqref{NC} with  
\begin{equation}
\mathcal{H}=\frac{1}{2}{\Phi'}^2+\frac{\nu}{2\beta^2}
e^{\beta\Phi}-\frac{2}{\beta}\Phi'',\quad 
\mathcal{P}=\Phi' \Pi-\frac{2}{\beta} \Pi'\label{HY3}.
\end{equation}
Their Poisson algebra is centrally extended, it is given by
\eqref{eq:2} where 
\begin{multline}
  \label{eq:9}
   K_{\xi_1,\xi_2}=-\frac{4}{\beta^2}\int_0^{2\pi}d\phi \big[T_1
   Y'''_2+Y_1T'''_2]\\=\frac{2}{\beta^2} \int_0^{2\pi}d\phi \big[T'_1 Y_2''+Y'_1 T''_2
  -(1\leftrightarrow 2)\big]. 
\end{multline}
In terms of modes defined again by \eqref{eq:10}, one gets the
centrally extended \(\mathfrak{bms}_3\) algebra \eqref{bms3} with
\(c_1=0,\frac{c_2}{12}=\frac{8\pi}{\beta^2}\). Note that in this case,
we may see from (\ref{couplings}) that the constant $G$ is kept
finite, because $\beta=\sqrt{32\pi G}$ is held fixed in the limit. The
value of the central charge turns out to be precisely the
gravitational one.

\section{Energy-momenum tensor, B\"acklund transformation and general solution}
\label{sec:backl-gener-solut}

\subsection{Liouville theory}
\label{sec:liouville-theory-1}

In this section, we recall the classical part of the analysis in
\cite{Braaten:1982fr,Braaten:1982yn,DHoker:1982er} .

When using the Hamiltonian equations of motion, the charge densities
satisfy 
\begin{equation}
  \label{eq:23}
  \d_u \cH = \frac{1}{l^2} \d_\phi \cP,\quad 
\d_u \cP= \d_\phi \cH. 
\end{equation}
They are thus given by 
\begin{equation}
  \label{eq:24}
  \cH=\frac{4}{\gamma^2l^2}(\Xi_{++}+\Xi_{--}),\quad
  \cP=\frac{4}{\gamma^2l}(\Xi_{++}-\Xi_{--}),
\end{equation}
with \(\Xi_{++}=\Xi_{++}(x^+)\), \(\Xi_{--}=\Xi_{--}(x^-)\) and the
conserved charges reduce on-shell to 
\begin{equation}
  \label{eq:25}
  Q_\xi=\frac{4}{\gamma^2 l} \int^{2\pi}_0 d\phi\, (Y^+\Xi_{++}+
  Y^-\Xi_{--}),
\end{equation}
where the normalization is chosen here in order to agree with
conventions used in the gravitational context.  Equivalently, one can
first express the energy-momentum tensor in light-cone coordinates,
\begin{equation}
T_{\pm\pm}=\frac{l}{2}\big(l\cH\pm
  \cP\big)=\frac{1}{4} (l\pi\pm\varphi')^2 +\frac{\mu l^2}{4\gamma^2}
e^{\gamma\varphi} \mp\frac{1}{\gamma} ( l\pi\pm\varphi')',\label{eq:29}
\end{equation}
so that \(T_{\pm\pm}=\frac{4}{\gamma^2}
\Xi_{\pm\pm}\).  One recovers the more familiar form on-shell,
\begin{equation}
  \label{eq:36}
  T_{\pm\pm}= (\d_\pm \varphi)^2-\frac{2}{\gamma}\d_\pm^2
  \varphi,\quad T_{\pm\mp}=0.  
\end{equation}
Conservation is equivalent to \(\d_{\mp}
T_{\pm\pm}=0\) and the transformation laws follow from \eqref{eq:18},
\begin{equation}
  \label{eq:35}
  \tilde
T_{++}=(F')^{-2}\big(T_{++}+\frac{2}{\gamma^2}\{F;x^+\}\big),
\end{equation}
where \(\{F;x^+\}=\frac{F'''}{F'}-\frac{3}{2}\frac{(F'')^2}{(F')^2}=(\ln
F')''-\half ((\ln F')')^2\) denotes the Schwarzian derivative and similarly for
\(T_{--}\). 

Let us now assume \(\mu\geq 0\). The B\"acklund transformation from
Liouville theory to a free field \(\psi\) with momentum \(\pi_\psi\)
is the canonical transformation determined by 
\begin{equation}
\begin{split}
\label{bl}
\int_0^{2\pi}d\phi\,
\pi\dot{\varphi}-H[\varphi,\pi]=\int_0^{2\pi}d\phi\,
\pi_{\psi}\dot{\psi}-K[\psi,\pi_{\psi}]+ \frac{d}{du}W[\varphi,\psi],
\\
W[\varphi,\psi]=\int_0^{2\pi} d\phi \left[\frac{1}{l}\varphi \psi'-
  \frac{2 }{\gamma^2}
  \sqrt{\mu}e^{\frac{\gamma\varphi}{2}}\sinh(\frac{\gamma\psi}{2})\right].
\end{split}
\end{equation}
This gives the transformation equations
\begin{equation}
\begin{split}
  \pi=\frac{\delta W}{\delta \varphi}=\frac{1}{l}\psi'-
\frac{1}{\gamma}
\sqrt{\mu}e^{\frac{\gamma\varphi}{2}}\sinh\left(\frac{\gamma\psi}{2}\right),
 \label{backpi}\\
  \pi_{\psi}=-\frac{\delta W}{\delta
    \psi}=\frac{1}{l}\varphi'+\frac{1}{\gamma}
\sqrt{\mu}e^{\frac{\gamma\varphi}{2}}
\cosh\left(\frac{\gamma\psi}{2}\right). 
\end{split}
\end{equation}
When used in the integrand of \(H[\varphi,\pi]\) one finds, after an
integration by parts and another use of the last of relations
\eqref{backpi}, that
\begin{equation}
\label{K1}
K[\psi,\pi_\psi]=\int^{2\pi}_0d\phi\, \left(\frac{1}{2}\pi_{\psi}^2+\frac{1}{2
l^2}\psi'^2\right),
\end{equation}
which is the Hamiltonian of a free massless field in two dimensions. A
useful form for its solution is
\begin{equation}
\label{solsK}
\psi=\frac{1}{\gamma
  }\ln\left(\frac{A'}{B'}\right),\quad
  \pi_{\psi}=\dot{\psi},\quad A=A(x^+),\ B=B(x^-).
\end{equation}
One may find
the general solution \(\varphi\) to
Liouville's equation by replacing  $\psi$ above in  the second of relations
\eqref{backpi},
 \begin{equation}
 \label{Liouvillesol}
e^{\gamma\varphi}=\frac{16}{ l^2 \mu
}\frac{A'B'}{(A-B)^2}=\frac{16}{l^2\mu}
\frac{C'B'}{(1+CB)^2}, \quad A=-\frac{1}{C}. 
\end{equation}

Finally, this expression can be used to express the energy-momentum
tensor in terms of the arbitrary functions appearing in the general solution, 
\begin{equation}
  T_{++}=-\frac{2}{\gamma^2}
  \{A;x^+\}=-\frac{2}{\gamma^2}\{C;x^+\},\quad  T_{--}=-\frac{2}{\gamma^2}
  \{B;x^-\}.
\end{equation}

\subsection{Non-centrally extended limit}
\label{sec:no-centr-extend}

On-shell, the charge densities, and thus the components of the
energy-momentum tensor, now satisfy \(\d_u\cH = 0\) and
\(\d_u \cP = \d_\phi \cH\), so that they are given by 
\begin{equation}
  \label{eq:19}
  \cH= \frac{2}{\sigma^2}\Theta,\quad
  \cP=\frac{2}{\sigma^2}(2\Xi+u\Theta'),\quad
  \Theta=\Theta(\phi),\ \Xi=\Xi(\phi),
\end{equation}
for some normalization \(\sigma\). On-shell, the charges reduce to 
\begin{equation}
  \label{eq:26}
  Q_\xi=\frac{2}{\sigma^2}\int^{2\pi}_0d\phi\,
  (T\Theta+2Y \Xi). 
\end{equation}
The on-shell transformation laws for the functions determining the
energy-momentum tensor can then be worked out and are given by
\begin{equation}
  \label{eq:37}
\begin{split}
&  \tilde\Theta(\tilde\phi)=(\tilde\phi')^{-2}\Theta,\\
& \tilde
\Xi(\tilde\phi)=(\tilde\phi')^{-2}\Big[\Xi-\frac{\alpha}{2}
\Theta'-\alpha' \Theta
\Big].
\end{split}
\end{equation}
The associated infinitesimal versions are 
\begin{equation}
  \label{eq:38}
\begin{split}
&  -\delta \Theta=Y\Theta' +2Y'\Theta,\\
& -\delta \Xi = Y\Xi'+2Y'\Xi +\half T \Theta' +T'\Theta. 
\end{split}
\end{equation}

In the B\"acklund transformations \eqref{bl},\eqref{backpi} and
\eqref{K1}, the terms proportional to \(l^{-1},l^{-2}\) drop out, so
that 
\begin{equation}
  \label{eq:30}
  K[\psi,\pi_\psi]=\int^{2\pi}_0d\phi\, \half \pi_\psi^2.
\end{equation}
The free field \(\psi\) now satisfies \(\ddot \psi=0\) and so is
given by
\begin{equation}
  \label{eq:27}
  \psi=\frac{1}{\gamma}(A +u  B),\quad 
 A= A(\phi),\  B= B(\phi). 
\end{equation}
The second equation of \eqref{backpi} now yields
\begin{equation}
  \label{eq:28}
  e^{\gamma\varphi}=\frac{ B^2}{\mu \cosh^2{\frac{
        A+u B}{2}}}. 
\end{equation}
Again, on-shell, the arbitrary functions determining the components of
the energy-momentum tensor can be expressed in terms of the
arbitrary functions appearing in the general solution,
\begin{equation}
{\Theta}=\frac{\sigma^2}{4\gamma^2}{B}^2, \quad
{\Xi}=\frac{\sigma^2}{4\gamma^2}{A}'{B}. 
\end{equation}

\subsection{Centrally extended limit}
\label{sec:centr-extend-limit-1}

The charge densities again satisfy \(\d_u\cH = 0\) and
\(\d_u \cP = \d_\phi \cH\) on-shell, so that 
\begin{equation}
  \label{eq:19a}
  \cH= \frac{2}{\beta^2}\Theta,\quad
  \cP=\frac{2}{\beta^2}(2\Xi+u\Theta'),\quad
  \Theta=\Theta(\phi),\ \Xi=\Xi(\phi),
\end{equation}
with on-shell charges given by 
\begin{equation}
  \label{eq:16}
  Q_\xi=\frac{2}{\beta^2}\int^{2\pi}_{0} (T\Theta +2 Y \Xi).
\end{equation}
The on-shell transformation laws for the functions determining the
energy-momentum tensor are now given by
\begin{equation}
  \label{eq:37b}
\begin{split}
&  \tilde\Theta(\tilde\phi)=(\tilde\phi')^{-2}\big[\Theta(\phi)+
2\{\tilde\phi;\phi\}\big],\\
& \tilde
\Xi(\tilde\phi)=(\tilde\phi')^{-2}\Big[\Xi-\frac{\alpha}{2}
\Theta'-\alpha' \Theta+\alpha'''
\Big].
\end{split}
\end{equation}
The associated infinitesimal versions are 
\begin{equation}
  \label{eq:38b}
\begin{split}
&  -\delta \Theta=Y\Theta' +2Y'\Theta-2Y''',\\
& -\delta \Xi = Y\Xi'+2Y'\Xi +\half T \Theta' +T'\Theta-T'''. 
\end{split}
\end{equation}
The normalization \(\frac{2}{\beta^2}\) chosen above is
conventional. The choice made here is such that the transformation
laws agree with the gravitational ones. In the latter context
\(\Theta,\Xi\) denote the arbitrary functions that appear in the
general solution to asymptotically flat gravity in three dimensions in
\({\rm BMS}\) gauge (cf.~section 3 of \cite{Barnich:2010eb}).

The B\"acklund transformations are now determined by 
\begin{equation}
\begin{split}
\label{bl2}
\int_0^{2\pi}d\phi\,
\Pi\dot{\Phi}-H[\Phi,\Pi]=\int_0^{2\pi}d\phi\,
\pi_{\psi}\dot{\psi}-K[\psi,\pi_{\psi}]+ \frac{d}{du}W[\Phi,\psi],
\\
W[\Phi,\psi]=\int_0^{2\pi} d\phi \left[\Phi \psi'-
  \frac{1}{\beta}
  \sqrt{\nu}e^{\frac{\beta\Phi}{2}}\psi\right],
\end{split}
\end{equation}
so that 
\begin{equation}
\begin{split}
  \Pi=\frac{\delta W}{\delta \varphi}=\psi'-
\frac{1}{2}
\sqrt{\nu}e^{\frac{\beta\Phi}{2}}\psi,
 \label{backpi2}\\
  \pi_{\psi}=-\frac{\delta W}{\delta
    \psi}=\Phi'+\frac{1}{\beta}
\sqrt{\nu}e^{\frac{\beta\Phi}{2}}. 
\end{split}
\end{equation}
This gives again
\begin{equation}
  \label{eq:30a}
  K[\psi,\pi_\psi]=\int^{2\pi}_0d\phi\, \half \pi_\psi^2.
\end{equation}
For the solution of the free theory, we now choose
\begin{equation}
  \label{eq:15}
  \psi=\frac{1}{\beta}(A + 2 u (\ln B')'),\quad 
  A= A(\phi),\  B= B(\phi). 
\end{equation}
and from \eqref{backpi2}, one then finds the local solution 
\begin{equation}
\begin{split}
  \label{eq:17}
  e^{\beta\Phi} & =\frac{4}{\nu}((\ln B)')^2 ,\\
\beta\Pi & =\frac{A'B-B'A}{B}+2u\left((\ln B')''-\frac{B''}{B}\right).
\end{split}
\end{equation}
  
In this case, the relation between the arbitrary functions in the
energy-momentum tensor and those in the general solution is 
\begin{equation}
\Theta=-2\{B;\phi\}, 
\quad \Xi=\frac{1}{2}\frac{A'B''-B'A''}{B'}.
\end{equation}
 
 \section{Conclusions}
 
 In this note, we have taken a short-cut for constructing an action
 describing the boundary degrees of freedom of (2+1)-dimensional,
 asymptotically flat Einstein gravity. In order to do so, we have
 taken appropriate ``flat'' limits of Liouville, which is known to be
 the theory that describes the boundary dynamics in the asymptotically
 anti-de Sitter case.  The limit may be taken in, at least, two
 different ways. Both give rise to \({\rm BMS}_3\) invariant
 two-dimensional field theories.  Whereas the first limit has no
 central extension, the second one admits a central extension of
 exactly the same type than in the gravitational surface charge
 algebra.

 The constructed theories are interacting two-dimensional field
 theories with a symmetry group, namely \({\rm BMS}_3\), that is of
 the same dimension than the conformal algebra.  We have explicitly
 constructed the finite symmetry transformations and constructed the
 conserved charges in each theory. As for conformally invariant
 theories, these charges are related to the corresponding
 energy-momentum tensors, which are also given explicitly. We have
 constructed the most general solutions of both theories making use of
 the B\"acklund transformations which, as for Liouville, maps the
 non-linear to a free field theory.

 We have worked in the canonical formulation. It turns out that for the case
 with vanishing central extension, the momentum may be eliminated in
 the Hamiltonian action principle, leading us to a second order,
 Lagrangian action. In the centrally extended case, which is the one
 appropriated for describing gravity, one cannot eliminate the
 momentum. One may, however, eliminate the original field in terms of
 the momentum. This gives rise to a spatially non-local Lagrangian.

 The complete analysis, which will be carried out in follow-up work,
 consists in starting from the first order Chern-Simons formulation of
 three dimensional gravity and implementing the Hamiltonian reduction
 required by the gravitational boundary conditions on the associated
 WZW theory to end up with the proposed centrally extended flat limit
 of Liouville theory. 

\section*{Acknowledgements}
\label{sec:acknowledgements}

\addcontentsline{toc}{section}{Acknowledgments}

The authors are grateful to M.~Ba\~nados, G.~Giribet, M.~Pino, C.~Troessaert
and R.~Troncoso for useful discussions.  This work is supported in part
by the Fund for Scientific Research-FNRS (Belgium), by IISN-Belgium,
by ``Communaut\'e fran\c caise de Belgique - Actions de Recherche
Concert\'ees'' and by Fondecyt Projects No.~1085322 and
No.~1090753. The work of A.G.~was partially supported by Fondecyt
(Chile) Grant \#1090753. H.G.~thanks Conicyt for financial support.


\begin{thebibliography}{10}

\bibitem{Seiberg:1990eb}
N.~Seiberg, ``{Notes on quantum Liouville theory and quantum gravity},'' {\em
  Prog. Theor. Phys. Suppl.} {\bf 102} (1990)
319--349.

\bibitem{Teschner:2001rv}
J.~Teschner, ``{Liouville theory revisited},'' {\em Class.Quant.Grav.} {\bf 18}
  (2001) R153--R222, \href{http://www.arXiv.org/abs/hep-th/0104158}{{\tt
  hep-th/0104158}}.

\bibitem{Nakayama:2004vk}
Y.~Nakayama, ``{Liouville field theory: A Decade after the revolution},'' {\em
  Int.J.Mod.Phys.} {\bf A19} (2004) 2771--2930,
  \href{http://www.arXiv.org/abs/hep-th/0402009}{{\tt hep-th/0402009}}.

\bibitem{Aharony:1999ti}
O.~Aharony, S.~S. Gubser, J.~Maldacena, H.~Ooguri, and Y.~Oz, ``Large {N} field
  theories, string theory and gravity,'' {\em Phys. Rept.} {\bf 323} (2000)
  183--386,
\href{http://arXiv.org/abs/hep-th/9905111}{{\tt hep-th/9905111}}.

\bibitem{Coussaert:1995zp}
O.~Coussaert, M.~Henneaux, and P.~van Driel, ``{The asymptotic dynamics of
  three-dimensional Einstein gravity with a negative cosmological constant},''
  {\em Class. Quant. Grav.} {\bf 12} (1995) 2961--2966,
  \href{http://www.arXiv.org/abs/gr-qc/9506019}{{\tt gr-qc/9506019}}.

\bibitem{Achucarro:1986vz}
A.~Achucarro and P.~K. Townsend, ``A {C}hern-{S}imons action for
  three-dimensional anti-de {S}itter supergravity theories,'' {\em Phys. Lett.}
  {\bf B180} (1986)
89.

\bibitem{Witten:1988hc}
E.~Witten, ``{(2+1)-dimensional Gravity As An Exactly Soluble System},'' {\em
  Nucl. Phys.} {\bf B311} (1988)
46.

\bibitem{PhysRevD.55.669}
A.~Ashtekar, J.~Bicak, and B.~G. Schmidt, ``Asymptotic structure of
  symmetry-reduced general relativity,'' {\em Phys. Rev. D} {\bf 55} (Jan,
  1997) 669--686.

\bibitem{Barnich:2006avcorr}
G.~Barnich and G.~Comp{\`e}re, ``Classical central extension for asymptotic
  symmetries at null infinity in three spacetime dimensions,'' {\em Class.
  Quant. Grav.} {\bf 24} (2007) F15,
  \href{http://www.arXiv.org/abs/gr-qc/0610130}{{\tt gr-qc/0610130}}.
Corrigendum: ibid 24 (2007) 3139.

\bibitem{Barnich:2010eb}
G.~Barnich and C.~Troessaert, ``{Aspects of the BMS/CFT correspondence},'' {\em
  JHEP} {\bf 05} (2010) 062,
\href{http://www.arXiv.org/abs/1001.1541}{{\tt 1001.1541}}.

\bibitem{Barnich:2012aw}
G.~Barnich, A.~Gomberoff, and H.~A. Gonz\'alez, ``{Flat limit of three
  dimensional asymptotically anti-de Sitter spacetimes},'' {\em Phys. Rev. D}
  {\bf 86} (2012) 024020,
\href{http://www.arXiv.org/abs/1204.3288}{{\tt 1204.3288}}.

\bibitem{Barnich:2012he}
G.~Barnich, ``{Entropy of three-dimensional asymptotically flat cosmological
  solutions},''
\href{http://www.arXiv.org/abs/1208.4371}{{\tt 1208.4371}}.

\bibitem{Bagchi:2012he}
A.~Bagchi, S.~Detournay, R.~Fareghbal, and J.~Simon, ``{Holography of 3d Flat
  Cosmological Horizons},''
\href{http://www.arXiv.org/abs/1208.4372}{{\tt 1208.4372}}.

\bibitem{Strominger:1998eq}
A.~Strominger, ``Black hole entropy from near-horizon microstates,'' {\em JHEP}
  {\bf 02} (1998) 009,
\href{http://www.arXiv.org/abs/arXiv:hep-th/9712251}{{\tt
  arXiv:hep-th/9712251}}.

\bibitem{Brown:1986nw}
J.~D. Brown and M.~Henneaux, ``Central charges in the canonical realization of
  asymptotic symmetries: An example from three-dimensional gravity,'' {\em
  Commun. Math. Phys.} {\bf 104} (1986) 207.

\bibitem{Braaten:1982fr}
E.~Braaten, T.~Curtright, and C.~B. Thorn, ``{Quantum Backlund Transformation
  for the Liouville Theory},'' {\em Phys. Lett.} {\bf B118} (1982)
115.

\bibitem{Braaten:1982yn}
E.~Braaten, T.~Curtright, and C.~B. Thorn, ``{An Exact Operator Solution of the
  Quantum Liouville Field Theory},'' {\em Ann. Phys.} {\bf 147} (1983)
365.

\bibitem{DHoker:1982er}
E.~D'Hoker and R.~Jackiw, ``{Liouville Field Theory},'' {\em Phys. Rev.} {\bf
  D26} (1982)
3517.

\end{thebibliography}

\def\cprime{$'$}
\providecommand{\href}[2]{#2}\begingroup\raggedright\endgroup

\end{document}